\documentclass[fleqn,10pt]{SelfArx} 

\definecolor{color1}{RGB}{0,0,90} 
\definecolor{color2}{RGB}{0,20,20} 

\usepackage{hyperref} 
\hypersetup{hidelinks,colorlinks,breaklinks=true,urlcolor=color2,citecolor=color1,linkcolor=color1,bookmarksopen=false,pdftitle={Title},pdfauthor={Author},urlcolor=blue}

\usepackage{graphicx}
\usepackage[square]{natbib}
\usepackage{amsmath}
\usepackage{multirow}
\usepackage{subfigure}
\usepackage{epstopdf}

\newcommand{\n}[1]{\mathrm{#1}}

\newcounter{tempEquationCounter}
\newcounter{thisEquationNumber}
\newenvironment{floatEq}
{\setcounter{thisEquationNumber}{\value{equation}}\addtocounter{equation}{1}
\begin{figure*}[!t]
\normalsize\setcounter{tempEquationCounter}{\value{equation}}
\setcounter{equation}{\value{thisEquationNumber}}
}
{\setcounter{equation}{\value{tempEquationCounter}}
\hrulefill\vspace*{4pt}
\end{figure*}

}

\JournalInfo{Published in Journal of Applied Physics, Vol. 114 (5), 053912, 2013} 
\Archive{\href{http://dx.doi.org/10.1063/1.4816096}{DOI: 10.1063/1.4816096}} 

\PaperTitle{Metamaterial anisotropic flux concentrators and magnetic arrays} 

\Authors{R. Bj\o{}rk, A. Smith and C. R. H. Bahl} 
\affiliation{\textit{Department of Energy Conversion and Storage, Technical University of Denmark - DTU, Frederiksborgvej 399, DK-4000 Roskilde, Denmark}} 
\affiliation{*\textbf{Corresponding author}: rabj@dtu.dk} 

\Keywords{} 
\newcommand{\keywordname}{Keywords} 

\Abstract{A metamaterial magnetic flux concentrator is investigated in detail in combination with a Halbach cylinder of infinite length. A general analytical solution to the field is determined and the magnetic figure of merit is determined for a Halbach cylinder with a flux concentrator. It is shown that an ideal flux concentrator will not change the figure of merit of a given magnet design, while the non-ideal will always lower it. The geometric parameters producing maximum figure of merit, i.e. the most efficient devices, are determined. The force and torque between two concentric Halbach cylinders with flux concentrators is determined and the maximum torque is found. Finally, the effect of non-ideal flux concentrators and the practical use of flux concentrators, as well as demagnetization issues, is discussed.}

\begin{document}

\flushbottom 

\maketitle 


\thispagestyle{empty} 

\section{Introduction}
Magnetic metamaterials, i.e.\ materials that are artificially modified at a length scale larger than the atomic, have been the subject of increasing interest. A special case is that of metamaterials which can modify static magnetic fields, i.e. fields at zero frequency. One example is a magnetic cloak, which when placed in an external magnetic field leaves that field unchanged while an interior cavity in the cloak experiences zero field. Such a material was suggested by Wood and Pendry \cite{Wood_2007} (and others) and investigated experimentally by Magnus et al.\cite{Magnus_2008} and G\"om\"ory et al.\cite{Gomory_2012}

However, metamaterials may be used to actively modify the field instead of used for shielding purposes. Recently, Navau et al.\cite{Navau_2012} suggested an interesting metamaterial object consisting of a cylindrical ring (or a spherical shell) made of a material with an anisotropic, spatially constant magnetic permeability ($\mu_r \neq \mu_\phi$, where $\mu_r$ and $\mu_\phi$ are the relative permeabilities in the radial and tangential directions, respectively). If $\mu_r > \mu_\phi$ this object tends to concentrate external flux lines in the space which it encloses; we will henceforth refer to it as a flux concentrator. Navau et al.\ show that in the limit $\mu_{r} \rightarrow \infty{}$ and $\mu_{\phi}\rightarrow{}0$ while keeping $\mu_r\mu_\phi = 1$ such a flux concentrator will leave a uniform external field unchanged on the outside of the concentrator while generating a uniform field inside it equal to the external field multiplied by a factor of $R_m/R_i$ where $R_m$ and $R_i$ are the outer and inner radius of the flux concentrator, respectively. The focus of Navau et al.\ was the use of such a structure for harvesting and redistributing magnetic energy in space. However, one can also consider the effect of a flux concentrator in conjunction with a permanent magnet configuration. Then the object is not to harvest magnetic energy but to change the field produced by the magnet configuration in a given volume. The object of the present paper is to analyze the field produced by the combination of a permanent magnet configuration with a flux concentrator.

Permanent magnet constructions that produce a powerful magnetic flux density, either uniformly or with a prescribed spatial variation, in a specified volume are used in a number of applications, such as nuclear magnetic resonance (NMR) equipment \cite{Moresi_2003,Appelt_2006}, accelerator magnets \cite{Sullivan_1998,Lim_2005} and magnetic refrigeration devices \cite{Tura_2007,Bjoerk_2010b}. A widely used permanent magnet configuration is the so-called Halbach cylinder which can generate either a uniform flux density or a multipole field. In the limit where the length of the array is much larger than its diameter, the field from the magnet can to a good approximation be found by considering a Halbach cylinder of infinite length.. This two-dimensional problem becomes amenable to analytical calculations. The Halbach cylinder is a hollow cylinder made of a ferromagnetic material with a remanent flux density which in cylindrical coordinates is given by
\begin{align}
B_{\mathrm{rem},r}    &= B_{\mathrm{rem}}\; \cos p\phi \nonumber\\
B_{\mathrm{rem},\phi} &= B_{\mathrm{rem}}\; \sin p\phi,\label{Eq.Halbach_magnetization}\tag{1}
\end{align}
where $B_{\mathrm{rem}}$ is the magnitude of the remanent flux density and $p$ is an integer \cite{Mallinson_1973,Halbach_1980}. For $p$ positive an internal field is generated, which for the important case of $p=1$ is spatially uniform. Outside the cylinder, the field is identically zero. For $p$ negative the Halbach cylinder creates a field on its outside, while the inside field becomes zero. The magnetic field distribution for a Halbach cylinder of infinite length \cite{Zhu_1993,Atallah_1997,Peng_2003,Xia_2004,Bjoerk_2010a} as well as for finite length \cite{Mhiochain_1999,Xu_2004,Bjoerk_2008,Bjoerk_2011b} have previously been investigated in detail. However, the question of the field from a Halbach cylinder with a flux concentrator does not seem to have been considered before in the literature.

\begin{floatEq}
\begin{equation}\label{interiorfield}
  \begin{pmatrix} B_r(r,\phi) \\ B_\phi(r,\phi) \end{pmatrix} = \begin{cases} B_\n{rem}\frac{p}{p-1}\left(\frac{R_i}{R_m}\right)^{\kappa p-1}\left[1- \left(\frac{R_m}{R_o}\right)^{p-1}\right]\frac{4\lambda}{(1+\lambda)^2-(1-\lambda)^2(R_i/R_m)^{2\kappa p}}\left(\frac{r}{R_i}\right)^{p-1} \left(\begin{smallmatrix} \cos p\phi \\ -\sin p\phi \end{smallmatrix}\right) & p > 1 \\
  B_\n{rem}\left(\frac{R_i}{R_m}\right)^{\kappa-1}\ln\left(\frac{R_o}{R_m}\right) \frac{4\lambda}{(1+\lambda)^2-(1-\lambda)^2(R_i/R_m)^{2\kappa}}
  \left(\begin{smallmatrix} \cos \phi \\ -\sin \phi \end{smallmatrix}\right) & p=1. \end{cases}\tag{2}
\end{equation}
\end{floatEq}

The plan of the paper is as follows: First we calculate the field generated by a $p$-Halbach in combination with a flux concentrator analytically and discuss how the flux concentrator impacts the field. The figure of merit, i.e. efficiency, of such a system, considered as a device for the generation of a magnetic field, is then determined. We then discuss the case of two concentric Halbachs together with two flux concentrators and show how the magnetic field as well as the torque between the inner and outer Halbach is influenced by the flux concentrators. Materials with infinite permeability in one direction and zero permeability in another do not exist, of course. Therefore we examine some possible realizations of such as a material as a composite metamaterial. Finally, we discuss the implications of our findings.

\section{Analytical results}
\begin{figure}[!t]
  \centering
  \includegraphics[width=1\columnwidth]{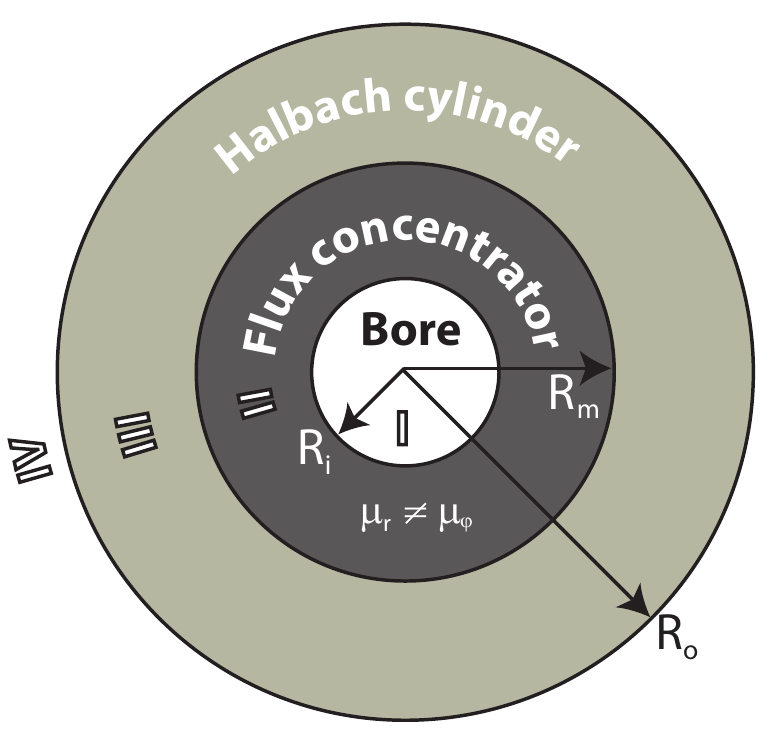}
  \caption{The combined Halbach cylinder and flux concentrator system for a $p$-Halbach ($p>0$). The different radii and regions have been indicated.}
  \label{Fig_Combined_drawing}
\end{figure}

Let us consider the field from a $p$-Halbach in combination with a flux concentrator with radial permeability $\mu_r$ and tangential permeability $\mu_\phi$; for the present we make no assumptions regarding their relative sizes. The geometry is as shown in Fig.~\ref{Fig_Combined_drawing} for the case of an `interior' Halbach (positive $p$); for `exterior' Halbachs (corresponding to negative $p$) the flux concentrator is placed concentrically on the outside of the cylinder. We assume that the permanent magnets are perfectly linear, i.e.\ with an infinite intrinsic coercivity.

\begin{floatEq}
\begin{equation}\label{MstarFlux}
M = \begin{cases} \left(\frac{R_i}{R_m}\right)^{2\kappa p}\left(\frac{4\lambda}{(1+\lambda)^2-(1-\lambda)^2(R_i/R_m)^{2\kappa p}}\right)^{2} \frac{p}{(1-p)^2}\left(1-\left(\frac{R_m}{R_o}\right)^{p-1}\right)^2 \frac{(R_m/R_o)^2}{1-(R_m/R_o)^2} & p > 1 \\
  \left(\frac{R_i}{R_m}\right)^{2\kappa p}\left(\frac{4\lambda}{(1+\lambda)^2-(1-\lambda)^2(R_i/R_m)^{2\kappa p}}\right)^{2} \ln\left(\frac{R_o}{R_m}\right)^2 \frac{(R_m/R_o)^2}{1-(R_m/R_o)^2} & p=1 \\
  -\left(\frac{R_O}{R_o}\right)^{2\kappa p}\left(\frac{4\lambda}{(1+\lambda)^2-(1-\lambda)^2(R_O/R_o)^{2\kappa p}}\right)^2 \frac{(R_m/R_o)^2}{1-(R_m/R_o)^2}\left(\frac{R_m}{R_o}\right)^{-2p}\frac{p}{(1-p)^2}\left(1-\left(\frac{R_m}{R_o}\right)^{p-1}\right)^2 & p\leq -1. \end{cases}\tag{7}
\end{equation}
\end{floatEq}

To find the field, we introduce the vector potential $A_z(r,\phi)$ through $\mathbf{B}= \nabla\times (0,0,A_z)$. We then solve the Maxwell equation $\nabla\times\mathbf{H} = 0$ in each of the domains, subject to the boundary conditions of continuity of $B_r$ and $H_\phi$ (see Appendix~\ref{analytics} for details). The results for the field may be expressed in terms of the two parameters $\kappa = \sqrt{\mu_\phi/\mu_r}$ and $\lambda = \sqrt{\mu_r\mu_\phi}$. For the interior Halbach we get Eq. (\ref{interiorfield}), where $R_o$ is the outer radius of the Halbach, $R_m$ the inner radius (equal to the outer radius of the flux concentrator), while $R_i$ is the inner radius of the flux concentrator. For $p=1$ the generated field is still uniform, while for $p>1$ the field has the same spatial and angular dependence as for a $p$-Halbach without the flux concentrator. Thus, in both cases the effect of the flux concentrator is to multiply the field by a constant factor, allowing us to summarize the results as
\begin{equation}\label{field}
  \begin{split}
    \mathbf{B}^I(r,\phi) & = \left(\frac{R_i}{R_m}\right)^{\kappa p - p}\\ & \quad\cdot\frac{4\lambda}{(1+\lambda)^2-(1-\lambda)^2(R_i/R_m)^{2\kappa p}} \mathbf{B}_0^I(r,\phi).
\end{split}\tag{3}
\end{equation}
where $\mathbf{B}^I$ denotes the flux density in region I in Fig. \ref{Fig_Combined_drawing} and $\mathbf{B}_0^I$ denotes the field in the same region, in the case without a flux concentrator. However, it should be noted that if $\lambda \neq 1$ there will now be a field on the outside of the cylinder, in contrast to the case without the concentrator. This field can be found from Appendix~\ref{analytics}.

For an exterior Halbach ($p\leq -1$) we get for the field outside the Halbach and flux concentrator ($R_m$ and $R_o$ are the inner and outer radius of the Halbach; $R_o$ and $R_O$ are the inner and outer radius of the flux concentrator):
\begin{equation}\label{exteriorfield}
  \begin{split}\begin{pmatrix} B_r(r,\phi) \\ B_\phi(r,\phi) \end{pmatrix} & = B_\n{rem}\frac{p}{p-1}\left(\frac{R_O}{R_o}\right)^{\kappa p-1}\left[1- \left(\frac{R_o}{R_m}\right)^{p-1}\right] \\ & \quad\cdot\frac{4\lambda}{(1+\lambda)^2-(1-\lambda)^2(R_O/R_o)^{2\kappa p}}\\ & \quad\cdot\left(\frac{R_O}{r}\right)^{-p+1} \begin{pmatrix} \cos p\phi \\ -\sin p\phi \end{pmatrix}.\end{split}\tag{4}
\end{equation}

The resulting flux density, $\mathbf{B}^{IV}$, is also a constant times the field without the flux concentrator, $\mathbf{B}_0^{IV}$:
\begin{equation}\label{fieldneg}
  \begin{split} \mathbf{B}^{IV}(r,\phi) & = \left(\frac{R_O}{R_o}\right)^{\kappa p - p} \\ & \quad\cdot\frac{4\lambda}{(1+\lambda)^2-(1-\lambda)^2(R_O/R_o)^{2\kappa p}} \mathbf{B}_0^{IV}(r,\phi).\end{split}\tag{5}
\end{equation}

In this case there is also a field inside the Halbach if $\lambda\neq 1$. Both for an interior and an exterior Halbach the field generated is the same for $\lambda$ and $1/\lambda$. The $\lambda$-dependent factor is in both cases positive and bounded by 1.

For the case of $\lambda = 1$ the above expressions simplify considerably and the field outside the interior Halbachs and inside the exterior Halbachs becomes identically zero, in generalization of the result found by Navau et al.\cite{Navau_2012} If we compare now the two extreme cases $\kappa = 1$ (an ordinary isotropic material, i.e.\ no flux concentration) and $\kappa = 0$ (maximally anisotropic flux concentrator) we find that the effect of the flux concentrator is to multiply the field generated by a $p$-Halbach by the simple factor $(R_m/R_i)^p$ (interior Halbach) or $(R_O/R_o)^{-p}$ (exterior Halbach). This has implications both for the figure of merit of a given combination of Halbach cylinder and flux concentrator and for the force and torque experienced by concentric Halbachs. These questions will be addressed below.

Finally, we note that if $\kappa > 1$, i.e.\ if $\mu_\phi > \mu_r$, the generated field decreases by a factor of $(R_m/R_i)^{\kappa p}$ (for an interior Halbach) compared to the isotropic case. Thus, in this case the flux concentrator acts as a `flux diluter'.

\subsection{Figure of merit of a Halbach cylinder with a flux concentrator}
The object of a permanent magnet array is to generate a magnetic field of given characteristics in a given volume. Many different magnet configurations can in principle produce the same magnetic field, and thus the question arises of how to do it most efficiently. Jensen and Abele\cite{Jensen_1996} proposed a general figure of merit, $M$, to characterize the efficiency of a given magnet design:
\begin{equation}\label{Eq.Mstar_definition}
M=\frac{\int_{V_\n{field}}||\mathbf{B}||^2dV}{\int_{V_\n{mag}}||\mathbf{B_\n{rem}}||^2dV}=\frac{\int_{V_\n{mag}}(-\mathbf{B}\cdot{}\mathbf{H})dV}{4\mu_r\int_{V_\n{mag}}(-\mathbf{B}\cdot{}\mathbf{H})_{\textrm{max}}dV},\tag{6}
\end{equation}
where $V_\n{field}$ is the volume of the region where the magnetic field is created and $V_\n{mag}$ is the volume of the magnets. The figure of merit is the ratio of the energy stored in the field region to the maximum amount of magnetic energy available in the magnetic material, or formulated in terms of the permanent magnet energy product ($-\mathbf{B}\cdot{}\mathbf{H})$, the average energy product to the average of the maximum energy product. It can be shown that the maximum value of $M$ is 0.25 \cite{Jensen_1996}. Thus, this figure of merit parameter measures how well utilized the magnets are, when generating a specific magnetic field.

The general magnetic figure of merit of a Halbach cylinder have been calculated \cite{Bjoerk_2013b}, but here we consider the combined system with a flux concentrator. When a flux concentrator is added to the Halbach cylinder the domain of integration of the field in Eq.~(\ref{Eq.Mstar_definition}) changes from the edges of the Halbach cylinder to those of the flux concentrator. The flux density is given by Eqs.~(\ref{field}) and (\ref{fieldneg}). The figure of merit is given in Eq. (\ref{MstarFlux}).

The expressions in Eq.~(\ref{MstarFlux}) have been verified by comparison with simulation results from Comsol Multiphysics. For $\kappa=0$ and $\lambda=1$ the figure of merit becomes equal to the figure of merit of a Halbach cylinder without flux concentrator \cite{Bjoerk_2010a}, while for $\kappa > 0$ or $\lambda\neq 1$ the figure of merit will decrease. Thus only in the case of an ideal flux concentrator, the figure of merit of the magnet system will be unchanged when a flux concentrator is added to the system. This is due to the fact that in the interior of the ideal flux concentrator $\mathbf{B}\cdot\mathbf{H}=0$, i.e.\ no magnetic energy is stored inside it. The maximum figure of merit will still occur for the same ratio of the inner and outer radius as for the imperfect flux concentrator. Since this will be true for any given magnet design adding an ideal flux concentrator will not change the figure of merit of a magnet design. However, Eq. (\ref{MstarFlux}) can be used to calculate the figure of merit for a system with an imperfect flux concentrator, which is useful for practical applications. Furthermore, because the figure of merit at best remains unchanged when an flux concentrator is used, certainly does not mean that the flux concentrator cannot be advantageous for certain applications. This will be considered in Sec. \ref{Sec.Applications}.

It is also of interest to consider a magnet efficiency parameters that differs from $M$, as specific applications might have a different field dependence than $B^2$. Consider e.g. the efficiency parameter for magnetic refrigeration for a completely uniform field, $\Lambda_\n{cool} = (B^{2/3}-B_\n{low}^{2/3})\frac{V_\n{field}}{V_\n{mag}}P_\n{field}$, where $B_\n{low}$ is the flux density in the low field region and $P_\n{field}$ is the fraction of time magnetocaloric material is in the field \cite{Bjoerk_2008,Bjoerk_2010b}. This efficiency parameter is proportional to $B^{2/3}$ as the adiabatic temperature change due to the magnetocaloric effect scales with this value close to the Curie temperature. Adding a magnetic flux concentrator to a system will increase $B$, but also decrease $V_\n{field}$. Performing the integration one can see that if the relevant figure of merit is proportional to $B^{\alpha{}}$ for $\alpha{}<2$ the efficiency of a design will always decrease when adding a flux concentrator, while it will increase for $\alpha > 2$, assuming that the efficiency is directly proportional to $V_\n{mag}$. Thus for the $\Lambda_\n{cool}$ figure of merit a flux concentrator will always decrease the efficiency of a given design.

\subsection{Halbach with flux concentrator for $p=1$}\label{Effip1}
It is of interest to determine the optimal geometrical dimensions of the most efficient combined Halbach cylinder and flux concentrator system. For a Halbach cylinder without flux concentrator the optimal ratio of the radii does not have a closed form solution for $p>3$ \cite{Bjoerk_2010a}, and this will also be the case for a system with flux concentrator. We therefore consider the most common Halbach cylinder, namely the $p=1$ Halbach cylinder in combination with a flux concentrator. We wish to determine the optimal geometrical size of the flux concentrator for a given desired value of $B$ and inner radius $R_\n{i}$. For now, we only consider a perfect flux concentrator, i.e. $\kappa{}=0$ and $\lambda{}=1$. Calculating and differentiating the expression for $B$ from Eq.~(\ref{interiorfield}), i.e. Eq. (\ref{Eq.A36}), with respect to $R_\n{m}$ and putting it equal to zero, one gets that the flux density will obtain its largest value when
\begin{equation}\label{Eq.Bmax_Rm_Ro}
R_\n{m} = R_\n{o}e^{-1},\tag{8}
\end{equation}
which will produce the maximum flux density, $B_\n{max}$, for a given inner and outer radius of the whole system, of
\begin{equation}\label{Eq.Bmax_Ro_Ri}
B_\n{max} = B_\n{rem}e^{-1}\frac{R_\n{o}}{R_\n{i}},\tag{9}
\end{equation}
with a corresponding magnet area of
\begin{align}\label{Eq.A_mag_for_Bmax}
A_\n{mag,B_\n{max}} &= \pi{}\left(e^{2}-1\right)\frac{B_\n{max}^2}{B_\n{rem}^2}R_\n{i}^2.\tag{10}
\end{align}

However, in order to obtain the maximum efficiency of a combined $p=1$ Halbach and flux concentrator the configuration that produces a desired flux density in a desired bore using the least amount of magnet material must be found. The cross-sectional area of the magnet, i.e. the Halbach cylinder, is given by
\begin{equation}\label{Eq.Vmag_def}
A_\n{mag} = \pi{}R_\n{m}^2\left(e^{2\frac{B}{B_\n{rem}}\frac{R_\n{i}}{R_\n{m}}}-1\right).\tag{11}
\end{equation}

The minimum amount of magnet material as function of $R_\n{m}$ can be found by differentiating $A_\n{mag}$ with respect to $R_\n{m}$ and setting the derivative equal to zero. The solution to this equation is
\begin{equation}\label{Rm_Vmag_min}
R_\n{m} = \frac{B}{B_\n{rem}}\frac{2R_\n{i}}{W(-2e^{-2})+2}\approx 1.2550\frac{B}{B_\n{rem}}R_\n{i},\tag{12}
\end{equation}
where $W$ is the Lambert W function. The argument of the Lambert W function is greater than $-1/e$, which means that the function is single-valued. Defining the constant $\omega=2/(W(-2e^{-2})+2)\approx{}1.2550$, the corresponding minimum amount of magnet is
\begin{equation}\label{Eq.A_mag_min}
A_\n{mag,min} = \pi \omega^2 (e^{\frac{2}{\omega}}-1)\frac{B^2}{B_\n{rem}^2}R_\n{i}^2,\tag{13}
\end{equation}
which is needed to produce a given flux density for a given inner radius for a combined Halbach cylinder and flux concentrator system. The corresponding values of $R_\n{o}$ can be found from isolating $R_o$ in the equation for the norm of B, Eq. (\ref{Eq.A36}), derived from Eq.~(\ref{interiorfield}) while the value for $R_m$ is given in Eq.~(\ref{Rm_Vmag_min}). The minimum cross-sectional area can be compared with the cross-sectional area obtained when using the maximum value of $B$, i.e. Eq.~(\ref{Eq.A_mag_for_Bmax}). The difference between these is a constant factor of
\begin{equation}
\frac{A_\n{mag,min}}{A_\n{mag,B_\n{max}}} = \frac{\omega^2 (e^{\frac{2}{\omega}}-1)}{e^{2}-1} \approx  0.967.\tag{14}
\end{equation}
The difference in value for $R_\n{m}$ is also a constant factor of
\begin{equation}
\frac{R_\n{m, min\; mag}}{R_\n{m, B_\n{max}}} = \omega.\tag{15}
\end{equation}
For $R_\n{o}$ the ratio depends on the ratio between $B$ and $B_\n{rem}$.

For the two systems producing either the maximum flux or the configuration with the least magnet material, the figure of merits becomes independent of geometry. Using Eqs.~(\ref{Eq.A_mag_for_Bmax}) and (\ref{Eq.A_mag_min}) the figure of merits become
\begin{align}
M_\n{B_\n{max}} &= \frac{1}{e^2-1}\approx{}0.157\nonumber\\
M_\n{A_\n{min}} &= \frac{1}{\omega^2 (e^{\frac{2}{\omega}}-1)}\approx{}0.162~,\tag{16}
\end{align}
for the two cases, respectively. This means that a combined Halbach cylinder and flux concentrator system can be designed whose figure of merit does not depend on the ratio of $B$ and $B_\n{rem}$, opposite the case for a $p=1$ Halbach without a flux concentrator. The reason for this is that a magnetic flux concentrator is simply used to increase the magnetic flux density of an already maximally efficient Halbach cylinder system without a flux concentrator to the desired value of $B$. This would be an ideal application for a flux concentrator. While the figure of merit of the overall systems remains the same as for a case without a flux concentrator, the field achieved is much greater than would otherwise be the case. This clearly illustrates the usefulness of flux concentrators.

\section{Concentric Halbachs with flux concentrators}
Two ordinary Halbach cylinders (without flux concentrators) which are placed concentrically one within the other exert a force and a torque on each other for certain values of their pole number. Both the force and the torque can be calculated from the Maxwell stress tensor. Thus, the force per unit length exerted by the outer cylinder on the inner is given in Cartesian coordinates by
\begin{equation}
\mathbf{F} = \oint_S \mathbb{T}\cdot\mathbf{n}ds,\tag{17}
\end{equation}
where the integration is done over a closed surface (i.e. a line in 2D) enclosing the inner but not the outer Halbach; it can conveniently by taken as a circle in the middle of the gap between the two Halbachs. The Maxwell stress tensor $\mathbb{T}$ has the components
\begin{equation}
  \begin{pmatrix} T_{xx} & T_{xy} \\ T_{yx} & T_{yy} \end{pmatrix} = \begin{pmatrix} \frac{1}{2}(B_x^2-B_y^2) & B_xB_y \\ B_xB_y & \frac{1}{2}(B_y^2-B_x^2) \end{pmatrix},\tag{18}
\end{equation}
while $\mathbf{n}$ is the outwards-directed normal to the integration surface.

It can be shown\cite{Bjoerk_2010a} that the force is only non-zero for $p_1 = 1- p_2$ and $p_2 > 1$ where the innermost cylinder is a $p_1$-Halbach, while the outer is a $p_2$-Halbach. In this case the force per unit length acting on the inner cylinder due to the field of the outer is
\begin{equation}\label{forcenoconcentrator}
  \begin{pmatrix} F_x \\ F_y \end{pmatrix} = \frac{2\pi}{\mu_0}K \begin{pmatrix} \cos p_1\phi_0 \\ \sin p_1\phi_0 \end{pmatrix},\tag{19}
\end{equation}
where we have assumed that the inner magnet is rotated an angle $\phi_0$ with respect to the outer; the positive constant $K$ is $K= B_\n{rem,1}B_\n{rem,2}(R_{m,2}^{p_1}-R_{o,2}^{p_1})(R_{o,1}^{p_2}-R_{m,1}^{p_2})$.

Similarly, there is only a torque between two ordinary Halbachs if $p \equiv p_2 = -p_1 > 0$. For $p>1$ the torque per unit length is
\begin{equation}\label{torquenoconp}
  \tau = \frac{2\pi}{\mu_0}\frac{p^2}{1-p^2}K_\tau\sin p_2\phi_0,\tag{20}
\end{equation}
where $K_\tau = B_\n{rem,1}B_\n{rem,2}(R_{m,2}^{1-p}-R_{o,2}^{1-p})(R_{o,1}^{p+1}-R_{m,1}^{p+1})$. The case $p=1$ has to be considered separately, and one gets
\begin{equation}\label{torquenocon1}
  \tau = -\frac{\pi}{\mu_0}K_\tau' \sin \phi_0,\tag{21}
\end{equation}
with $K_\tau' = B_\n{rem,1}B_\n{rem,2}(R_{o,1}^{2}-R_{m,1}^{2})\ln\frac{R_{o,2}}{R_{m,2}}$.

\begin{figure}[t]
  \centering
  \includegraphics[width=1\columnwidth]{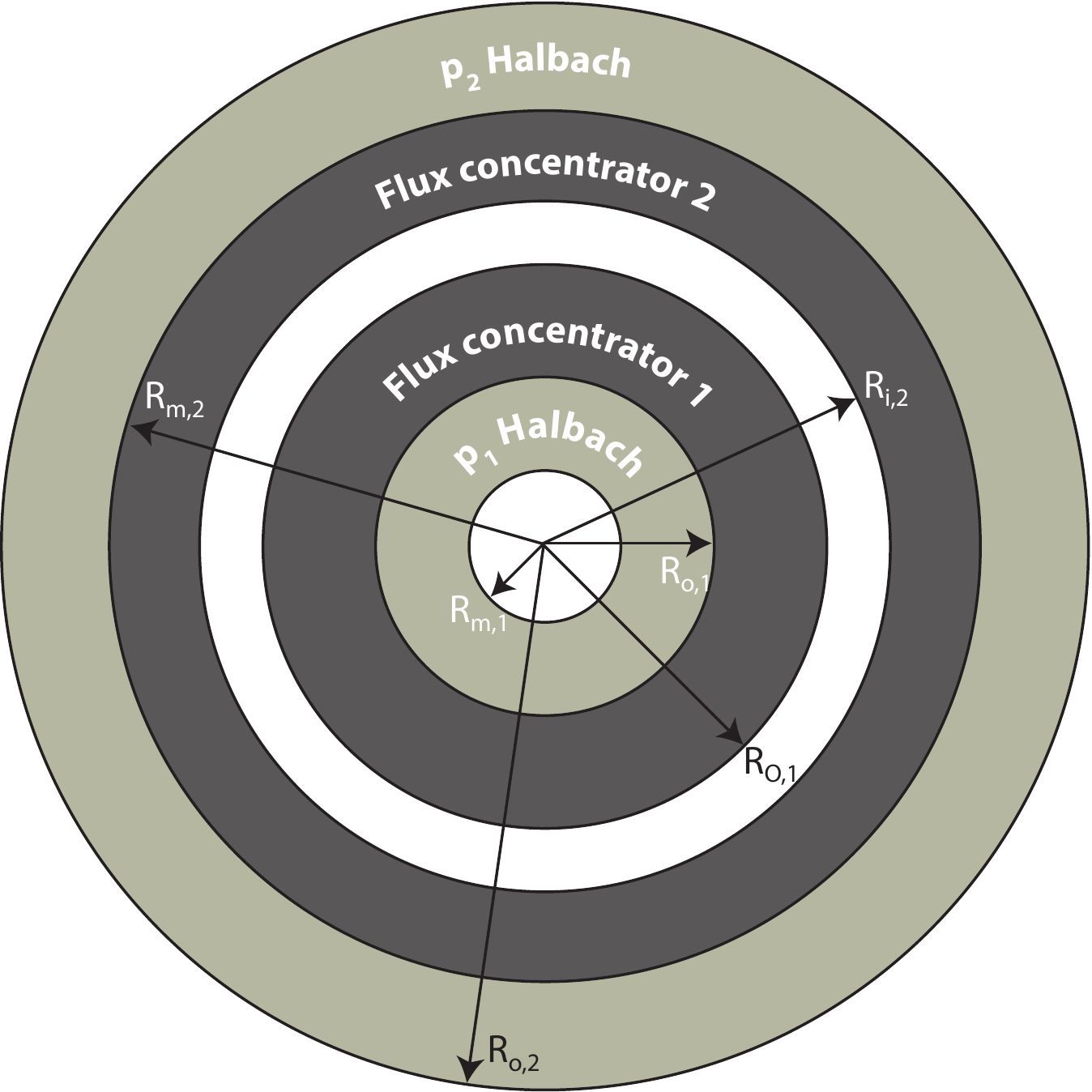}
  \caption{The two Halbachs, each equipped with a flux concentrator, used for the torque calculation. The different radii and regions have been indicated.}
  \label{Fig_Torque_drawing}
\end{figure}

If we now consider two Halbachs, each equipped with a flux concentrator (see Fig. \ref{Fig_Torque_drawing}), the force and the torque is modified. As the angular dependence of the Halbach fields is the same with and without a flux concentrator, it is still the case that the force is only non-zero for $p_1 = 1- p_2$ and $p_2 > 1$. Using equations \ref{field} and \ref{fieldneg} we immediately get that the force per unit length becomes
\begin{equation}\label{forcewithconcentrator}
  \begin{pmatrix} F_x \\ F_y \end{pmatrix} = \frac{2\pi}{\mu_0}K_\n{fc} \begin{pmatrix} \cos p_1\phi_0 \\ \sin p_1\phi_0 \end{pmatrix},\tag{22}
\end{equation}
where
\begin{equation}\label{kfc}
\begin{split}
  K_\n{fc} & = B_\n{rem,1}B_\n{rem,2}(R_{m,2}^{p_1}-R_{o,2}^{p_1})(R_{o,1}^{p_2}-R_{m,1}^{p_2}) \\ & \quad\cdot\left(\frac{R_{i,2}}{R_{m,2}}\right)^{\kappa_2 p_2 - p_2}\frac{4\lambda_2}{(1+\lambda_2)^2-(1-\lambda_2)^2(R_{i,2}/R_{m,2})^{2\kappa_2 p_2}} \\ & \quad\cdot\left(\frac{R_{O,1}}{R_{o,1}}\right)^{\kappa_1 p_1 - p_1}\frac{4\lambda_1}{(1+\lambda_1)^2-(1-\lambda_1)^2(R_{O,1}/R_{o,1})^{2\kappa_1 p_1}}.
\end{split}\tag{23}
\end{equation}
As before, there is only a torque for $p \equiv p_2 = -p_1 > 0$. The torque per unit length is
\begin{equation}\label{torquewithconcentrator}
  \tau = \begin{cases} \frac{p^2}{1-p^2}K_{\n{fc,}\tau}\sin p\phi_0 & p>1 \\
  -\frac{\pi}{\mu_0}K_{\n{fc,}\tau}' \sin \phi_0, & p=1 \end{cases}\tag{24}
\end{equation}
where
\begin{equation}\label{kfctau}
\begin{split}
  K_{\n{fc,}\tau} & = B_\n{rem,1}B_\n{rem,2}(R_{m,2}^{1-p}-R_{o,2}^{1-p})(R_{o,1}^{p+1}-R_{m,1}^{p+1}) \\
  & \quad\cdot\left(\frac{R_{i,2}}{R_{m,2}}\right)^{\kappa_2 p - p}\frac{4\lambda_2}{(1+\lambda_2)^2-(1-\lambda_2)^2(R_{i,2}/R_{m,2})^{2\kappa_2 p}} \\ & \quad\cdot\left(\frac{R_{O,1}}{R_{o,1}}\right)^{-\kappa_1 p + p}\frac{4\lambda_1}{(1+\lambda_1)^2-(1-\lambda_1)^2(R_{O,1}/R_{o,1})^{-2\kappa_1 p}}
\end{split}\tag{25}
\end{equation}
and
\begin{equation}\label{kfctauprime}
\begin{split}
  K_{\n{fc,}\tau}' & = B_\n{rem,1}B_\n{rem,2}(R_{o,1}^{2}-R_{m,1}^{2})\ln\frac{R_{o,2}}{R_{m,2}} \\
  & \quad\cdot\left(\frac{R_{i,2}}{R_{m,2}}\right)^{\kappa_2-1}\frac{4\lambda_2}{(1+\lambda_2)^2-(1-\lambda_2)^2(R_{i,2}/R_{m,2})^{2\kappa_2}} \\
  & \quad\cdot\left(\frac{R_{O,1}}{R_{o,1}}\right)^{-\kappa_1 +1}\frac{4\lambda_1}{(1+\lambda_1)^2-(1-\lambda_1)^2(R_{O,1}/R_{o,1})^{-2\kappa_1}}.
\end{split}\tag{26}
\end{equation}

With the technology of torque transfer through magnetic couplings in mind one may ask whether there is an optimal combination of $R_{m,1} \leq R_{o,1} \leq R_{O,1} \leq R_{i,2} \leq R_{m,2} \leq R_{o,2}$ which for a given outer radius $R_{o,2}$ will maximize the torque between the two Halbachs. It turns out (see Appendix~\ref{maxtorque}) that the maximum torque is achieved by having the two magnets fill the entire volume, i.e. with no flux concentrators and $R_{m,1}=0$, and with an outer radius of the inner magnet equal to the inner radius of the outer magnet equal to $R_\n{max}$, where
\begin{equation}\label{maxR}
  R_\n{max} = \begin{cases} e^{-1/2}R_{o,2} & p=1 \\ \left(\frac{2}{p+1}\right)^\frac{1}{p-1}R_{o,2} & p>1. \end{cases}\tag{27}
\end{equation}
Thus, it is not possible to increase the maximum amount of torque between two Halbachs by using flux concentrators.

Even though the maximum amount of torque cannot be increased, it should be noted that for two given (non-optimal) concentric Halbachs with some gap space between them, it is possible to increase the torque between them by filling the gap completely or partially with an ideal flux concentrator.


%


\section{Application of flux concentrators} \label{Sec.Applications}

It is of interest to consider the performance of a imperfect flux concentrator for application purposes. We consider a $p=1$ Halbach cylinder with maximally efficient dimensions, i.e. $R_o/R_m = 2.22$, as this is most frequently used in applications \cite{Abele_1990,Coey_2003}. The flux generated in the bore by placing an imperfect flux concentrator inside the Halbach cylinder can be found from Eq. (\ref{field}). In the following we will consider a flux concentrator with a fixed value of $\mu_r = 10^5$.

\begin{figure}[t]
  \centering
  \includegraphics[width=1\columnwidth]{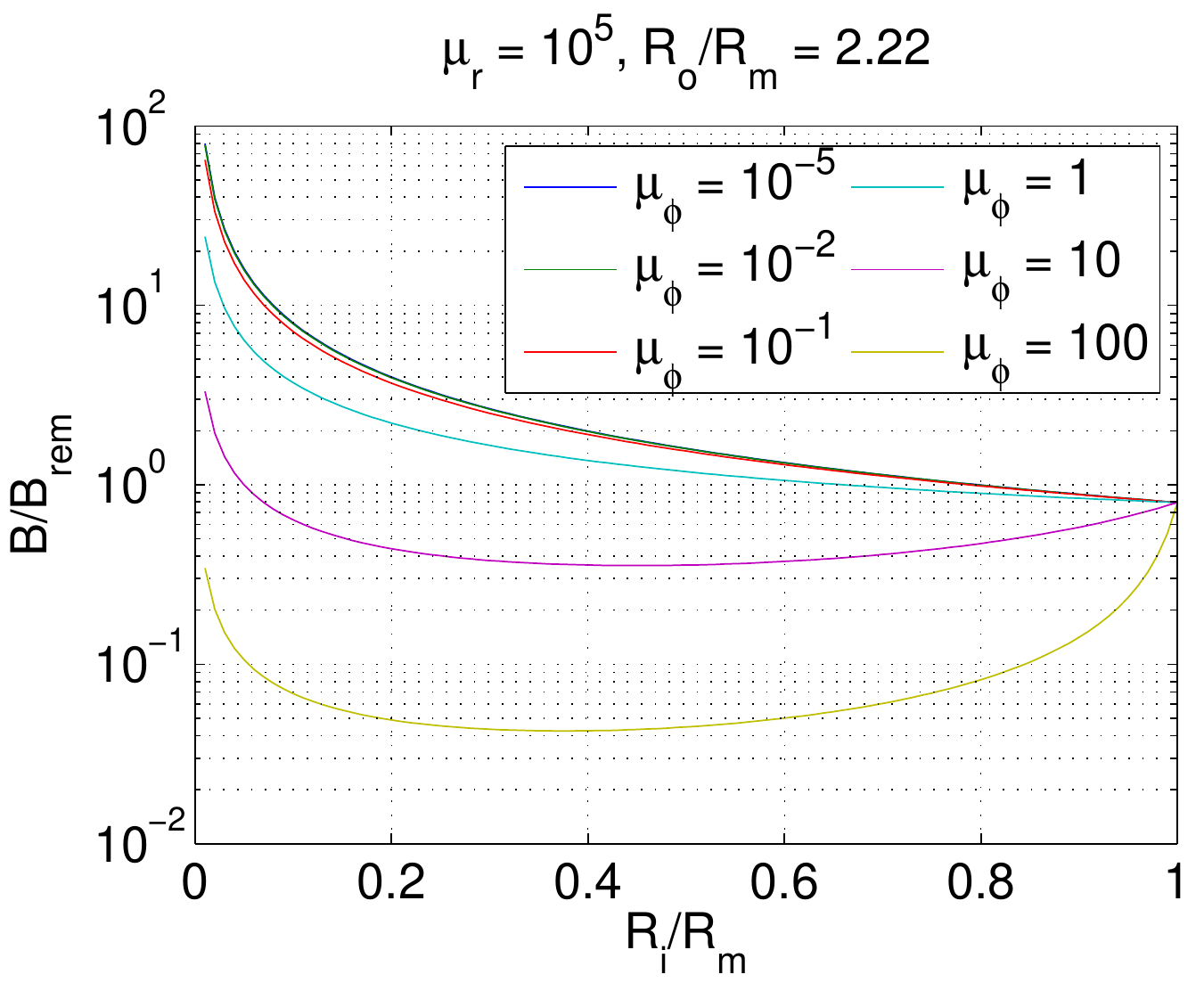}
  \caption{The normalized magnetic flux density as function of the size of the flux concentrator for different values of $\mu_\phi$. Note the logarithmic scale.}
  \label{Fig_B_p_1_lambda}
\end{figure}

\begin{figure}[t]
  \centering
  \includegraphics[width=1\columnwidth]{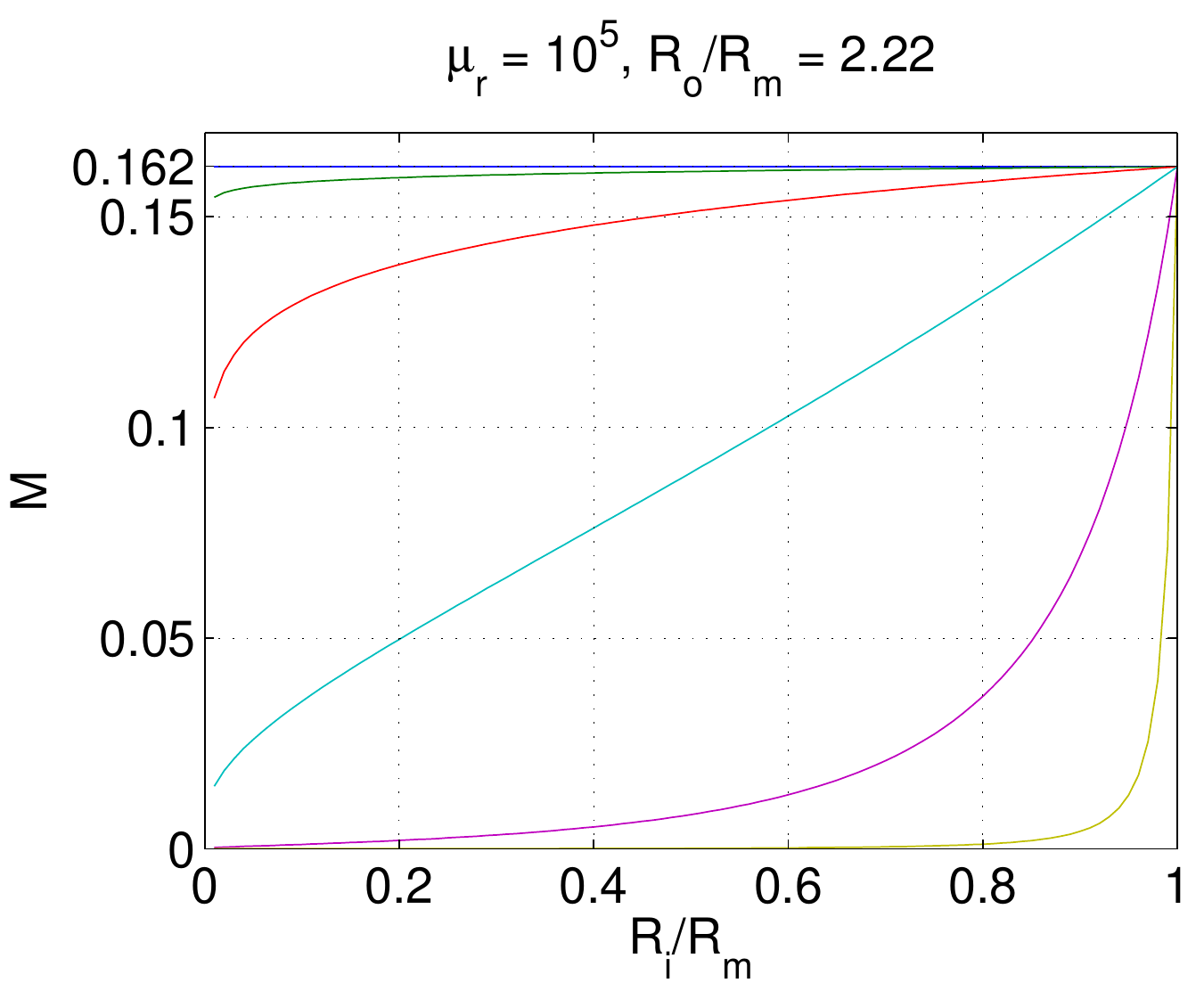}
  \caption{The figure of merit as function of the size of the flux concentrator for different values of $\mu_\phi$. The legend is identical to Fig. \ref{Fig_B_p_1_lambda}.}
  \label{Fig_Mstar_p_1_lambda}
\end{figure}

The flux density produced in the bore as function of the size of the flux concentrator is shown in Fig. \ref{Fig_B_p_1_lambda} for a range of values of $\mu_\phi$. It is clear that a high value of $\mu_\phi$ severely reduces the performance of the flux concentrator. This last point is better illustrated by showing the figure of merit of the combined system, as function of the size of the flux concentrator for the same range of $\mu_\phi$ values. This is shown in Fig. \ref{Fig_Mstar_p_1_lambda}. This figure clearly shows that increasing $\mu_\phi$ reduces the figure of merit of the flux concentrator severely.

A possible application for flux concentrators is for high field systems, for which demagnetization issues become relevant \cite{Kumada_2001a,Kumada_2001b,Kumada_2003}. Consider e.g. a $p=1$ Halbach for which the reverse component of the magnetic field can exceed the intrinsic coercivity in regions around the inner equator \cite{Bloch_1998, Bjoerk_2008}, as discussed previously. Using a combined Halbach cylinder and magnetic lens efficiently resolves this problem, as the Halbach cylinder in which the generated field does not exceed the coercivity can be used, and the magnetic lens can increase the field to the desired value, even though this is higher than the coercivity of the permanent magnets. This will be discussed subsequently.

\begin{figure}[t]
  \centering
  \includegraphics[width=1\columnwidth]{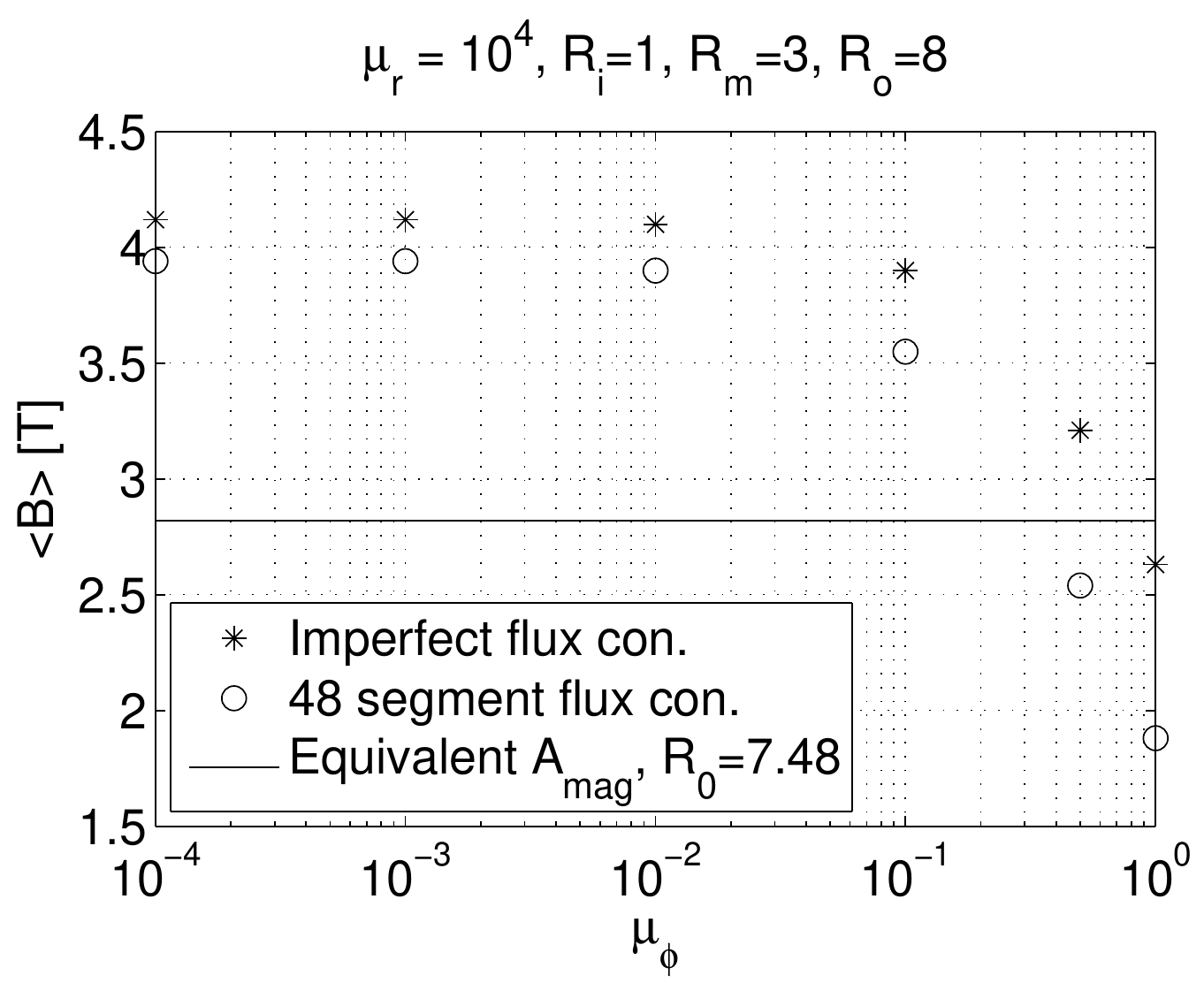}
  \caption{The average magnetic flux density in the bore for a $p=1$ Halbach with an imperfect flux concentrator as function of $\mu_\phi$ for the specific geometry. Also shown in the average field produced by the 48 segmented flux concentrator with alternating an alternating segments of materials with a isotropic permeability of $\mu_r$ and $\mu_\phi$, respectively. Finally the flux density produced by a Halbach alone with an equal amount of magnet, i.e. $R_o=7.48$ is shown.}
  \label{Fig_Anisotropic_and_48_segments}
\end{figure}

For application purposes it is also of interest to evaluate the performance of a segmented flux concentrator, as this might be a way to realize the flux concentrator design experimentally. Here we compare with the performance for an non-ideal but unsegmented flux concentrator. As an example we consider a system with $R_\n{i}=1$, $R_\n{m}=3$ and $R_\n{o}=8$ for a $p=1$ Halbach with a remanence of $B_\n{rem}=1.4$ T. Shown in Fig. \ref{Fig_Anisotropic_and_48_segments} is the average field in the bore for a non-ideal flux concentrator with $\mu_r=10^4$ and varying $\mu_\phi$ and for a 48 segmented flux concentrator with alternating segments of materials with a isotropic permeability of $\mu_r=10^4$ and $\mu_\phi$, respectively. Also shown is the the flux density produced by a Halbach alone with an equal amount of magnet, i.e. substituting the flux concentrator with magnet and changing the outer radius to $R_o=7.48$. As can be seen the 48 segmented design produces a consistently lower flux density than the non-segmented flux concentrator. In this example this means that for e.g. a value of $\mu_\phi=0.5$ a non-segmented flux concentrator still increases the flux density compared to a Halbach alone, whereas this is no longer the case if the flux concentrator is segmented in 48 parts.

\subsection{Demagnetization effects}
When a given Halbach design is implemented using permanent magnets with a finite coercivity, the demagnetization field internally in the magnet becomes an issue that need to be considered \cite{Katter_2005}. While the demagnetization of a $p$-Halbach cylinder alone has been considered elsewhere \cite{Bjoerk_2013b}, it is relevant to consider the influence of a flux concentrator on the demagnetization of a Halbach cylinder. The condition for demagnetization to occur is
\begin{equation}\label{demag}
  \mu_0\frac{\mathbf{H}\cdot\mathbf{M}}{B_\n{rem}} \leq -H_c.\tag{28}
\end{equation}
We consider the cases $p=1$ and $p>1$ separately. The case of $p<1$ is not considered, but the calculations follow the case of $p>1$.

For the case of $p=1$ the vector field inside the Halbach array is given by Eq.~(\ref{halbach1}). From this, the magnetic flux density is readily derived and using elementary trigonometric relations we get
\begin{equation}\label{demagp1}
  \mu_0\frac{\mathbf{H}\cdot\mathbf{M}}{B_\n{rem}} = \mu_0^{-1}B_\n{rem}\left((\ln\frac{R_o}{R_m}-\frac{1}{2})\cos 2\phi + \frac{K_2^{III}}{B_\n{rem}}r^{-2}-\frac{1}{2}\right),\tag{29}
\end{equation}
where $K_2^{III}$ is given by Eq.~(\ref{K2IIIp1}). The condition for demagnetization becomes
\begin{equation}\label{demagcondp1}
  -\frac{\mu_0H_c}{B_\n{rem}}+\frac{1}{2} > (\ln\frac{R_o}{r}-\frac{1}{2})\cos 2\phi + \frac{K_2^{III}}{B_\n{rem}}r^{-2}.\tag{30}
\end{equation}
These equations are identical to case for a Halbach cylinder without flux concentrator \cite{Bjoerk_2013b} except for the $K_2^{III}$ constant. This constant is equal to zero for the case without flux concentrator and also for the case for the conjugate flux concentrator, $\lambda=1$. For this case the field inside the Halbach is unchanged by the presence of the flux concentrator, regardless of the value of $\kappa$. In this case demagnetization will first arise for $\cos 2\phi=\pm1$.

For a $p>1$ Halbach cylinder the calculation of the demagnetization proceeds very similarly to the case of $p=1$. We get, using Eqs. (\ref{Halbachpgt1}) and (\ref{K1IIIpgt1}), that the condition for demagnetization to occur becomes
\begin{equation}\label{demagpgt1}
  -\frac{\mu_0H_c}{B_\n{rem}}+\frac{1}{2} > \left(-\frac{p}{p-1}\left(\frac{r}{R_o}\right)^{p-1}+\frac{1}{2}\frac{p+1}{p-1}\right)\cos 2p\phi + p\frac{K_2^{III}}{r^{p+1}}.\tag{31}
\end{equation}
Now, $K_2^{III}$ is given by Eq.~(\ref{K1IIIpgt1}). Again the equation is identical to the case without flux concentrator, except for the $K_2^{III}$ constant \cite{Bjoerk_2013b}. Thus the general conclusion is that flux concentrators with $\lambda>1$ tends to decrease demagnetization, while $\lambda<1$ increases it. A flux concentrator can thus be used to design a high flux density system, which would not have been possible without the flux concentrator due to demagnetization effects. Thus this constitutes another possible application for flux concentrators.

\section{Discussion and conclusion}
Magnetic metamaterials have long been studied theoretically, but their practical applications have been few. For the special case of metamaterials which can modify static magnetic fields, Navau et al. recently suggested a metamaterial object consisting of a hollow cylinder (or a spherical shell) made of a material with an anisotropic, spatially constant magnetic permeability. This object can concentrate external flux lines in the space which it encloses. By considering a well known cylindrical permanent magnetic structure, namely the Halbach cylinder, together with a flux concentrator, the influence of the flux concentrator on an actual magnetic structure can be investigated directly. The Halbach cylinder is a hollow cylinder made of a ferromagnetic material with a remanent flux density which is varied as a function of angle but remains constant in the radial direction.

Here we have calculated the field generated by a general Halbach cylinder in combination with a flux concentrator analytically. Having directly derived the analytical field equations allows for the magnetic efficiency of such a system to be determined directly and subsequently the optimal dimensions of such a system to be determined. It was shown that a flux concentrator cannot increase the figure of merit of a given magnet design. Following this the case of two concentric Halbachs together with two flux concentrators was discussed with emphasis on how the magnetic field as well as the torque between the inner and outer Halbach is influenced by the flux concentrators. The torque was calculated analytically and the maximum torque was determined. Finally, the possible realization of a metamaterial flux concentrator was discussed and the generated field and figure of merit of such constructs were discussed. It was showed that by constructing a segmented flux concentrator, the choice of materials and more specifically their relative permeability was critically important for the generated field. Demagnetization was also discussed and it was shown that flux concentrators may be useful for relieving demagnetization in permanent magnet constructs.

\appendix
\section*{Appendix}
\section{Analytical solution of the field}\label{analytics}
The field from a $p$-Halbach together with an anisotropic flux concentrator can readily be found analytically, using the methods outlined in, e.g., ref. \cite{Bjoerk_2010a}. For reference we briefly summarize the procedure here, together with the full results for the field distribution. A $p$-Halbach has the magnetization $\mu_0 M_r = B_\n{rem}\cos p\phi$, $\mu_0 M_\phi = B_\n{rem}\sin p\phi$ where $B_\n{rem}$ is the magnitude of the remanence. For $p>0$ the Halbach cylinder generates a field in the region inside it, while for $p<0$ it generates a field on the outside.

For the two-dimensional case considered here, we can write $\mathbf{B}=\nabla\times\mathbf{A} = \nabla\times (0,0,A_z)$. We solve the Maxwell equation $\nabla\times\mathbf{H} = 0$ in the Lorenz gauge $\nabla\cdot\mathbf{A}=0$. In cylindrical coordinates we get the following differential equation for $A_z(r,\phi)$ in the two vacuum regions:
\begin{equation}\label{i}
    r^2\frac{\partial^2A_z}{\partial r^2} + r\frac{\partial A_z}{\partial r}+\frac{\partial^2 A_z}{\partial \phi^2} = 0.\tag{A1}
\end{equation}
In the region containing the flux concentrator we have $H_r = \frac{1}{\mu_r\mu0} B_r$ and $H_\phi = \frac{1}{\mu_\phi\mu_0} B_\phi$ and get
\begin{equation}\label{iii}
    r^2\frac{\partial^2A_z}{\partial r^2} + r\frac{\partial A_z}{\partial r}+\kappa^2\frac{\partial^2 A_z}{\partial \phi^2} = 0,\tag{A2}
\end{equation}
where we have introduced $\kappa = \sqrt{\mu_\phi/\mu_r}$. Finally, inside the Halbach we have $H_r = \frac{1}{\mu_r\mu_0} B_r- M_r$ and $H_\phi = \frac{1}{\mu_\phi\mu_0} B_\phi- M_\phi$, and the equation to solve becomes
\begin{equation}\label{iv}
    r^2\frac{\partial^2A_z}{\partial r^2} + r\frac{\partial A_z}{\partial r}+\frac{\partial^2 A_z}{\partial \phi^2} = -B_\n{rem}(p+1)r \sin p\phi.\tag{A3}
\end{equation}
The solution has to be found subject to the conditions of continuity of $B_r$ and $H_\phi$ at each of the boundaries between the regions. This immediately constrains the general solution of the above equations to only contain terms with the angular dependence $\sin p\phi$.

\subsection{The interior Halbach with $p>1$}
We get the following solutions for the vector potential in each of the four regions:
\begin{align}\label{Halbachpgt1}
  A_z(r,\phi) &= K_1^Ir^p\sin p\phi \;\; \mathrm{(bore)} \nonumber\\
  A_z(r,\phi) &= (K_1^{II}r^{\kappa p}+K_2^{II}r^{-\kappa p})\sin p\phi  \;\; \mathrm{(flux\ concentrator)} \nonumber\\
  A_z(r,\phi) &= (K_1^{III}r^{p}+K_2^{III}r^{-p}+\frac{B_\n{rem}}{p-1}r)\sin p\phi  \;\; \mathrm{(Halbach)}  \nonumber\\
  A_z(r,\phi) &= K_2^{IV}r^{-p}\sin p\phi \;\; \mathrm{(outside)},\tag{A4}
\end{align}
where we have used the fact that the vector potential must remain finite for $r\rightarrow 0$ and $r\rightarrow \infty$ to set $K_2^I = K_1^{IV} = 0$.

The six constants are determined from the following six equations:
\begin{equation}
\begin{split}
 K_1^{I}R_i^{p-1} = K_1^{II}R_i^{\kappa p-1} + K_2^{II}R_i^{-\kappa p-1} \\
  K_1^{I}R_i^{p-1} = \lambda^{-1}(K_1^{II}R_i^{\kappa p-1} - K_2^{II}R_i^{-\kappa p-1}) \\
  K_1^{II}R_m^{\kappa p-1} + K_2^{II}R_m^{-\kappa p-1} = K_1^{III}R_m^{p-1} + K_2^{III}R_m^{-p-1} & \\ +\frac{1}{p-1}B_\n{rem} \\
  \lambda^{-1}(K_1^{II}R_m^{\kappa p-1} - K_2^{II}R_m^{-\kappa p-1}) = K_1^{III}R_m^{p-1} - K_2^{III}R_m^{-p-1} & \\+\frac{1}{p-1}B_\n{rem} \\
  K_1^{III}R_o^{p-1} + K_2^{III}R_o^{-p-1}+\frac{1}{p-1}B_\n{rem} = K_2^{IV}R_o^{-p-1} \\
  K_1^{III}R_o^{p-1} - K_2^{III}R_o^{-p-1}+\frac{1}{p-1}B_\n{rem} = -K_2^{IV}R_o^{-p-1}.
\end{split}
\tag{A5}
\end{equation}
Here, we have introduced $\lambda = \sqrt{\mu_r\mu_\phi}$.

This linear set of equations is straightforward to solve using, e.g., Mathematica. We get:
\begin{equation} \label{K1IIIpgt1}
\begin{split}
K_1^{I} & =  \frac{B_\n{rem}}{p-1}R_i^{-p+1}\left(\frac{R_i}{R_m}\right)^{\kappa p-1}\left[1- \left(\frac{R_m}{R_o}\right)^{p-1}\right]  \\ & \quad\cdot \frac{4\lambda}{(1+\lambda)^2-(1-\lambda)^2(R_i/R_m)^{2\kappa p}} \\
K_1^{II} & =  \frac{B_\n{rem}}{p-1}R_m^{-\kappa p+1}\left[1- \left(\frac{R_m}{R_o}\right)^{p-1}\right]  \\ & \quad\cdot \frac{2\lambda(1+\lambda)}{(1+\lambda)^2-(1-\lambda)^2(R_i/R_m)^{2\kappa p}} \\
K_2^{II} & =  \frac{B_\n{rem}}{p-1}R_m^{\kappa p+1}\left(\frac{R_i}{R_m}\right)^{2\kappa p}\left[1- \left(\frac{R_m}{R_o}\right)^{p-1}\right]  \\ & \quad\cdot \frac{2\lambda(1-\lambda)}{(1+\lambda)^2-(1-\lambda)^2(R_i/R_m)^{2\kappa p}} \\
K_1^{III} & = -\frac{B_\n{rem}}{p-1}R_o^{-p+1} \\
K_2^{III} & = \frac{B_\n{rem}}{p-1}R_m^{p+1}\left[1-\left(\frac{R_i}{R_m}\right)^{2\kappa p}\right]\left[1- \left(\frac{R_m}{R_o}\right)^{p-1}\right]  \\ & \quad\cdot \frac{\lambda^2-1}{(1+\lambda)^2-(1-\lambda)^2(R_i/R_m)^{2\kappa p}} \\
K_2^{IV} & =  K_2^{III}
\end{split}
\tag{A6}
\end{equation}
The field inside the bore is
\begin{equation}
\begin{split}
B_r^I (r,\phi) & = B_\n{rem}\frac{p}{p-1}\left(\frac{R_i}{R_m}\right)^{\kappa p-1}\left[1- \left(\frac{R_m}{R_o}\right)^{p-1}\right] \\ & \quad\cdot \frac{4\lambda}{(1+\lambda)^2-(1-\lambda)^2(R_i/R_m)^{2\kappa p}}\left(\frac{r}{R_i}\right)^{p-1}\cos p\phi \\
B_\phi^I (r,\phi) & = -B_\n{rem}\frac{p}{p-1}\left(\frac{R_i}{R_m}\right)^{\kappa p-1}\left[1- \left(\frac{R_m}{R_o}\right)^{p-1}\right] \\ & \quad\cdot \frac{4\lambda}{(1+\lambda)^2-(1-\lambda)^2(R_i/R_m)^{2\kappa p}}\left(\frac{r}{R_i}\right)^{p-1}\sin p\phi.
\end{split}
\tag{A7}
\end{equation}

In the absence of a flux concentrator, the $p>1$ Halbach cylinder generates the following field:\cite{Bjoerk_2010a}
\begin{equation}
\begin{split}
  B_{0,r}^{I} &= B_\n{rem}\frac{p}{p-1}\left[ 1-\left(\frac{R_m}{R_o}\right)^{p-1}\right]\left(\frac{r}{R_m}\right)^{p-1} \cos p\phi \nonumber\\
  B_{0,\phi}^I &= - B_\n{rem}\frac{p}{p-1}\left[ 1-\left(\frac{R_m}{R_o}\right)^{p-1}\right]\left(\frac{r}{R_m}\right)^{p-1} \sin p\phi.
\end{split}\tag{A8}
\end{equation}
Thus, the effect of the flux concentrator is to modify the interior field by a constant factor:
\begin{equation}\label{fcfield}
\begin{split}
    \mathbf{B}^I(r,\phi) & = \left(\frac{R_i}{R_m}\right)^{\kappa p - p} \\ & \quad\cdot \frac{4\lambda}{(1+\lambda)^2-(1-\lambda)^2(R_i/R_m)^{2\kappa p}} \mathbf{B}_0^I(r,\phi)
\end{split}
\tag{A9}
\end{equation}

If $\lambda = 1$, i.e.\ the radial and the tangential components of the permeability are \textit{conjugate} in the sense that $\mu_r=\mu_\phi^{-1}$, the above expressions simplify considerably. In that case the field in the bore becomes
\begin{equation}
\begin{split}
B_r^I (r,\phi) & = B_\n{rem}\frac{p}{p-1}\left(\frac{R_i}{R_m}\right)^{\kappa p-1}\left[1- \left(\frac{R_m}{R_o}\right)^{p-1}\right] \\ & \quad\cdot \left(\frac{r}{R_i}\right)^{p-1}\cos p\phi \\
B_\phi^I (r,\phi) & = -B_\n{rem}\frac{p}{p-1}\left(\frac{R_i}{R_m}\right)^{\kappa p-1}\left[1- \left(\frac{R_m}{R_o}\right)^{p-1}\right] \\ & \quad\cdot \left(\frac{r}{R_i}\right)^{p-1}\sin p\phi,
\end{split}
\tag{A10}
\end{equation}
and the field outside the Halbach becomes identically zero.

\subsection{The interior Halbach with $p=1$}
The case of $p=1$ has to be considered separately. The vector potential inside the Halbach now becomes
\begin{equation}\label{halbach1}
    A_z(r,\phi) = (K_1^{III}r+K_2^{III}r^{-1}-B_\n{rem}r\ln r)\sin \phi \;\; \mathrm{(Halbach)}.\tag{A11}
\end{equation}
Again we get six equations with six unknowns. The solution is
\begin{equation}
\begin{split}
K_1^{I} & =  B_\n{rem}\left(\frac{R_i}{R_m}\right)^{\kappa-1}\ln\left(\frac{R_o}{R_m}\right)  \\ & \quad\cdot \frac{4\lambda}{(1+\lambda)^2-(1-\lambda)^2(R_i/R_m)^{2\kappa}} \\
K_1^{II} & = B_\n{rem}R_m^{-\kappa+1}\ln\left(\frac{R_o}{R_m}\right)  \\ & \quad\cdot  \frac{2\lambda(1+\lambda)}{(1+\lambda)^2-(1-\lambda)^2(R_i/R_m)^{2\kappa}} \\
K_2^{II} & = B_\n{rem}R_m^{\kappa +1}\left(\frac{R_i}{R_m}\right)^{2\kappa}\ln\left(\frac{R_o}{R_m}\right)  \\ & \quad\cdot \frac{2(1-\lambda)\lambda}{(1+\lambda)^2-(1-\lambda)^2(R_i/R_m)^{2\kappa}} \\
K_1^{III} & = B_\n{rem}\ln R_o \\
K_2^{III} & = B_\n{rem}R_m^2\left[1-\left(\frac{R_i}{R_m}\right)^{2\kappa}\right]\ln\left(\frac{R_o}{R_m}\right)  \\ & \quad\cdot \frac{\lambda^2-1}{(1+\lambda)^2-(1-\lambda)^2(R_i/R_m)^{2\kappa}} \label{K2IIIp1} \\
K_2^{IV} & =  K_2^{III}
\end{split}
\tag{A12}
\end{equation}
The field inside the bore is constant, with a magnitude equal to
\begin{equation}
B^I = B_\n{rem}\left(\frac{R_i}{R_m}\right)^{\kappa-1}\ln\left(\frac{R_o}{R_m}\right) \frac{4\lambda}{(1+\lambda)^2-(1-\lambda)^2(R_i/R_m)^{2\kappa}}.\tag{A13}
\end{equation}
The field for the $p=1$ Halbach without a flux concentrator is $B_0^I = B_\n{rem}\ln(R_o/R_m)$, and again the effect of the flux concentrator is to multiply the field by a constant factor:
\begin{equation}\label{fcfield1}
    B^I = \left(\frac{R_i}{R_m}\right)^{\kappa-1}\frac{4\lambda}{(1+\lambda)^2-(1-\lambda)^2(R_i/R_m)^{2\kappa}}B_0^I.\tag{A14}
\end{equation}

For the case of a conjugate material we get for the field inside the bore:
\begin{equation}\label{Eq.A36}
B^I = B_\n{rem}\left(\frac{R_i}{R_m}\right)^{\kappa-1}\ln\left(\frac{R_o}{R_m}\right),\tag{A15}
\end{equation}
with the field outside the Halbach equal to zero.

\subsection{The exterior Halbach with $p\leq -1$}
The negative $p$ Halbach generates a field in the region outside it, while the field inside it is zero. Thus, it makes most sense to place the flux concentrator on the outside of the Halbach. We continue to call the interior and exterior radii of the Halbach for $R_m$ and $R_o$, but now the inner radius of the flux concentrator is $R_o$ while the outer radius is $R_O$.

For $p<-1$ we get the following solutions for the vector potential in each of the four regions:
\begin{equation}
\begin{split}
  A_z(r,\phi) &= K_2^Ir^{-p}\sin p\phi \;\; \mathrm{(bore)} \nonumber\\
  A_z(r,\phi) &= (K_1^{II}r^{p}+K_2^{II}r^{-p}+\frac{B_\n{rem}}{p-1}r)\sin p\phi \;\; \mathrm{(Halbach)} \nonumber\\
  A_z(r,\phi) &= (K_1^{III}r^{\kappa p}+K_2^{III}r^{-\kappa p})\sin p\phi \;\; \mathrm{(flux\ concentrator)} \nonumber\\
  A_z(r,\phi) &= K_1^{IV}r^{p}\sin p\phi \;\; \mathrm{(outside)},
\end{split}\tag{A16}
\end{equation}
subject to the six boundary conditions.

Again, the solution is straightforward, and we get:
\begin{equation}
\begin{split}
K_2^{I} & =  \frac{B_\n{rem}}{p-1}R_o^{p+1}\left[1-\left(\frac{R_O}{R_o}\right)^{2\kappa p}\right]\left[1-\left(\frac{R_o}{R_m}\right)^{p-1}\right] \\ & \quad\cdot \frac{\lambda^2-1}{(1+\lambda)^2-(1-\lambda)^2(R_O/R_o)^{2\kappa p}} \\
K_1^{II} & = -\frac{B_\n{rem}}{p-1}R_m^{-p+1} \\
K_2^{II} & =  K_2^{I} \\
K_1^{III} & = \frac{B_\n{rem}}{p-1}R_o^{-\kappa p+1}\left[1-\left(\frac{R_o}{R_m}\right)^{p-1}\right] \\ & \quad\cdot \frac{2\lambda(1+\lambda)}{(1+\lambda)^2-(1-\lambda)^2(R_O/R_o)^{2\kappa p}} \\
K_2^{III} & = \frac{B_\n{rem}}{p-1}R_o^{\kappa p+1}\left(\frac{R_O}{R_o}\right)^{2\kappa p}\left[1-\left(\frac{R_o}{R_m}\right)^{p-1}\right] \\ & \quad\cdot \frac{2\lambda(1-\lambda)}{(1+\lambda)^2-(1-\lambda)^2(R_O/R_o)^{2\kappa p}} \\
K_2^{IV} & =  \frac{B_\n{rem}}{p-1}R_O^{-p+1}\left(\frac{R_O}{R_o}\right)^{\kappa p-1}\left[1-\left(\frac{R_o}{R_m}\right)^{p-1}\right] \\ & \quad\cdot \frac{4\lambda}{(1+\lambda)^2-(1-\lambda)^2(R_O/R_o)^{2\kappa p}}.
\end{split}\tag{A17}
\end{equation}

The field outside the Halbach is then
\begin{equation}
\begin{split}
B_r^{IV} (r,\phi) & = B_\n{rem}\frac{p}{p-1}\left(\frac{R_O}{R_o}\right)^{\kappa p-1}\left[1- \left(\frac{R_o}{R_m}\right)^{p-1}\right] \\ & \quad\cdot \frac{4\lambda}{(1+\lambda)^2-(1-\lambda)^2(R_O/R_o)^{2\kappa p}} \\ & \quad\cdot \left(\frac{R_O}{r}\right)^{-p+1}\cos p\phi \\
B_\phi^{IV} (r,\phi) & = -B_\n{rem}\frac{p}{p-1}\left(\frac{R_O}{R_o}\right)^{\kappa p-1}\left[1- \left(\frac{R_o}{R_m}\right)^{p-1}\right] \\ & \quad\cdot \frac{4\lambda}{(1+\lambda)^2-(1-\lambda)^2(R_O/R_o)^{2\kappa p}} \\ & \quad\cdot \left(\frac{R_O}{r}\right)^{-p+1}\sin p\phi.
\end{split}\tag{A18}
\end{equation}
Compared to the case without a flux concentrator, the field is
\begin{equation}\label{fcfieldneg}
\begin{split}
    \mathbf{B}^{IV}(r,\phi) &= \left(\frac{R_O}{R_o}\right)^{\kappa p - p} \\ & \quad\cdot \frac{4\lambda}{(1+\lambda)^2-(1-\lambda)^2(R_O/R_o)^{2\kappa p}} \mathbf{B}_0^{IV}(r,\phi).
    \end{split}\tag{A19}
\end{equation}
Finally, the field in the conjugate case becomes
\begin{equation}
\begin{split}
B_r^{IV} (r,\phi) & = B_\n{rem}\frac{p}{p-1}\left(\frac{R_O}{R_o}\right)^{\kappa p-1}\left[1- \left(\frac{R_o}{R_m}\right)^{p-1}\right] \\ & \quad\cdot \left(\frac{R_O}{r}\right)^{-p+1}\cos p\phi \\
B_\phi^{IV} (r,\phi) & = -B_\n{rem}\frac{p}{p-1}\left(\frac{R_O}{R_o}\right)^{\kappa p-1}\left[1- \left(\frac{R_o}{R_m}\right)^{p-1}\right] \\ & \quad\cdot \left(\frac{R_O}{r}\right)^{-p+1}\sin p\phi.
\end{split}\tag{A20}
\end{equation}
The field inside the bore is identically zero.

For $p=-1$ the vector potential in region II is $A_z(r,\phi) = (K_1^{II}r^{p}+K_2^{II}r^{-p})\sin p\phi$. However, it turns out that the expressions for the six constants given above are still valid. In particular, the field outside the magnet is still given by Eq.~(\ref{fcfieldneg}).

\section{Maximum torque between two concentric Halbach arrays}\label{maxtorque}
The torque experienced by a $-p$-Halbach placed inside a $p$-Halbach is given by Eq.~(\ref{torquewithconcentrator}). The cases $p=1$ and $p>1$ have to be considered separately. However, in both cases it is evident that the torque considered as a function of the flux concentrator parameters $\lambda_1,\kappa_1,\lambda_2,\kappa_2$ attains its maximum for $\lambda_1=\lambda_2=1$ and $\kappa_1=\kappa_2=0$. Thus, we only need to consider these values.
\subsection{The maximum torque for $p=1$}
The magnitude of the torque is a constant times
\begin{equation}
  (R_{o,1}^2-R_{m,1}^2)\ln\left(\frac{R_{o,2}}{R_{m,2}}\right)\left(\frac{R_{i,2}}{R_{m,2}}\right)^{-1}\left(\frac{R_{O,1}}{R_{o,1}}\right).\tag{B1}
\end{equation}
We keep the outer radius of the outer Halbach fixed and put it equal to 1, i.e.\ measure all radii in units of $R_{o,2}$. Then the object is to maximize
\begin{equation}
  f(x_1,x_2,x_3,x_4,x_5) = -(x_2^2-x_1^2)(\ln x_5) x_4 x_5^{-1} x_3 x_2^{-1}\tag{B2}
\end{equation}
subject to the constraints $0 \leq x_1 \leq x_2 \leq x_3 \leq x_4 \leq x_5 \leq 1$. Since $\partial f/\partial x_1 < 0$ we can put $x_1 = 0$. Then all the partial derivatives $\partial f/\partial x_i$ are negative for $i=2,3,4$, i.e.\ $x_2$, $x_3$ and $x_4$ should be as large as possible. Thus we put $x_2=x_3= x_4 = x_5$ and the expression to be optimized becomes $-x_5^2 \ln x_5$. By equating its derivative to zero, the maximum is found to be at
\begin{equation}
  x_5 = e^{-1/2}.\tag{B3}
\end{equation}
\subsection{The maximum torque for $p>1$}
Now the expression to be maximized becomes
\begin{equation}
  g(x_1,x_2,x_3,x_4,x_5) = (x_5^{1-p}-1)(x_2^{p+1}-x_1^{p+1}) x_4^p x_5^{-p} x_3^p x_2^{-p}\tag{B4}
\end{equation}
subject to the constraints $0 \leq x_1 \leq x_2 \leq x_3 \leq x_4 \leq x_5 \leq 1$. Again we find $x_1=0$ and $x_2=x_3= x_4 = x_5$. Then we have to maximize $(x_5^{1-p}-1)x_5^{p+1}=x_5^2-x_5^{p+1}$ and get for the maximum that
\begin{equation}
  x_5 = \left(\frac{2}{p+1}\right)^{-\frac{1}{p-1}}.\tag{B5}
\end{equation}

%
%
%
%

\end{document}